\documentclass[conference,a4paper]{APSIPA2021}
\usepackage{multirow}
\usepackage{amsmath}
\usepackage[psamsfonts]{amssymb}
\usepackage{amsxtra}
\usepackage{threeparttable}

\usepackage[pdftex]{graphicx}
\usepackage{subfig}
\usepackage{url}
\usepackage{CJKutf8}
\newcommand{\ja}[1]{\begin{CJK}{UTF8}{min}#1\end{CJK}}
\newcommand{\blfootnote}[1]{\begingroup
\renewcommand\thefootnote{}\footnote{#1}%
\addtocounter{footnote}{-1}%
\endgroup}

\title{Prediction method of Soundscape Impressions using Environmental Sounds and Aerial Photographs}

\author{%
\authorblockN{%
	Yusuke Ono\authorrefmark{1},
	Sunao Hara\authorrefmark{1}\authorrefmark{2} and
	Masanobu Abe\authorrefmark{1}
}
\authorblockA{%
\authorrefmark{1}
Okayama University, Okayama, Japan}
\authorblockA{%
\authorrefmark{2}
E-mail: hara@okayama-u.ac.jp}
}

\begin{document}
\maketitle
\thispagestyle{empty}
\blfootnote{This work was supported by JSPS KAKENHI Grant Number JP20K12079.}

\begin{abstract}
We investigate an method for quantifying city characteristics based on impressions of a sound environment.
The quantification of the city characteristics will be beneficial to government policy planning, tourism projects, etc.
In this study, we try to predict two soundscape impressions, meaning pleasantness and eventfulness, using sound data collected by the cloud-sensing method.
The collected sounds comprise meta information of recording location using Global Positioning System.
Furthermore, the soundscape impressions and sound-source features are separately assigned to the cloud-sensing sounds by assessments defined using Swedish Soundscape-Quality Protocol, assessing the quality of the acoustic environment.
The prediction models are built using deep neural networks with multi-layer perceptron for the input of 10-second sound and the aerial photographs of its location.
An acoustic feature comprises equivalent noise level and outputs of octave-band filters every second, and statistics of them in 10~s.
An image feature is extracted from an aerial photograph using ResNet-50 and autoencoder architecture.
We perform comparison experiments to demonstrate the benefit of each feature.
As a result of the comparison, aerial photographs and sound-source features are efficient to predict impression information.
Additionally, even if the sound-source features are predicted using acoustic and image features, the features also show fine results to predict the soundscape impression close to the result of oracle sound-source features.
\end{abstract}

\section{Introduction}

Sounds in our daily lives are one of the most crucial information to detect the world around us.
Such crucial but unconscious sounds are called environmental sounds; e.g., people talking, car noise, birds singing, etc.
We can understand the impression or atmosphere of the location from environmental sounds that are frequently labeled as acoustic scenes, or soundscape.
The term ``soundscape'' defined in ISO 12913-1 \cite{ISO12913_1} is as follows:
\textit{acoustic environment as perceived or experienced and/or understood by a person or people, in context.}

Soundscapes are actively investigated recently \cite{axelsson10,lunden2016urban,pijanowski2011soundscape}.
In these studies, there are some proposals for protocols for collecting soundscape-related information.
For instance, Swedish Soundscape-Quality Protocol (SSQP) \cite{axelsson15,ISO12913_2} is one of the well-known assessment approaches for impressions of the sound environment.
The SSQP suggests, for instance, eight types of impression terms to denote soundscape impressions and taxonomy examples for sound sources.
It is based on a human annotation, however, if we could reduce the human process, we can efficiently use the large sound data collected using the cloud-sensing method;
e.g., EarPhone~\cite{Rana2010_IPSN_EarPhone}, NoiseSPY~\cite{Kanjo2010_NoiseSPY}, NoiseTube~\cite{DHondt2012_Journal_NoiseTube},
The Sound Around You Project~\cite{Mydlarz_InterNoise2011}, Otologmap~\cite{hara_ICME16,hara_APSIPA17}, etc.

In this study, we propose a prediction method for soundscape impression using acoustic features from environmental sounds and their meta information, e.g., image features from aerial photographs, and predicted sound-source features.
The labeling of soundscape impression and sound-source features were conducted on the environmental sound data~\cite{hara_APSIPA17}.
Deep neural network (DNN) models were built to predict the soundscape impressions and sound-source features.
As an experiment, we compared the accuracy of prediction that employed environmental sound, aerial photographs, and sound sources as input features, either alone or in combination.

\section{Soundscape impressions by SSQP} \label{chap:SSImpression}

\subsection{Related works}

One of the definitions for soundscape is published as an International Standard in ISO~12913-1~\cite{ISO12913_1}.
Data collection methods and reporting requirements for soundscape investigations are summarized as a Technical Specification (TS) in ISO/TS~12913-2~\cite{ISO12913_2}.
Furthermore, TS in ISO/TS~12913-3 described the methods for examining the collected data \cite{ISO12913_3}.

Axelsson~\cite{axelsson10} assessed what terms can be employed to describe environmental sounds. 
The participants listened to 30-second environmental sounds and rated them on a scale of 0 (not at all correct) to 100 (perfectly correct) using 116 various terms, respectively.
Then, he examined the findings with principal component analysis, and called the two principal components ``Pleasantness`` and ``Eventfulness.``
Lund{\'e}n and Hurtig~\cite{lunden2016urban} assessed 30-second environmental sounds based on these principal components.
They tackled training a machine learning model with support vector machine to predict the evaluation values from the acoustic features, mel-frequency cepstrum coefficients. 

From these studies, we assume that the predictor of soundscape impression could be built using acoustic features, which attain a certain level of prediction performance.
Furthermore, we focus on the lack of information, which should be experienced by an actual user, improving the performance.
The soundscape impressions should be also felt depending on visual information offered by the location's atmosphere, environmental objects around us, etc.
Thus, we tackle predicting the soundscape impressions with acoustic data and aerial photographs of the location in which the acoustic data is recorded.


\subsection{Soundscape attributes and impressions}

The Swedish Soundscape-Quality Protocol (SSQP) \cite{axelsson15} is suggested based on the work of \cite{axelsson10}.
It suggested assessment protocols for asking numerous questions about sound environments.
The question's detail is provided as a questionnaire method A in ISO/TS~12913-2~\cite{ISO12913_2}.

\begin{figure}[tb]
  \begin{center}
    \includegraphics{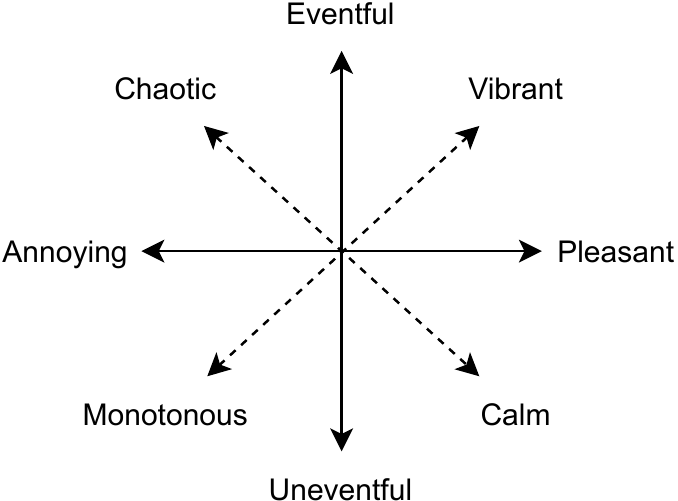}
    \caption{Graphical representation of soundscape attributes by SSQP \cite{ISO12913_3}}
    \label{fig:SSQP}
  \end{center}
\end{figure}

Figure~\ref{fig:SSQP} shows that impressions of the sound environments are represented by eight terms.
The eight terms are also called ``soundscape attributes.''
Each attribute is rated using a five-point scale from 1: Strongly disagree to 5: Strongly agree.

ISO/TS~12913-3 \cite{ISO12913_3} offers a two-dimensional model for summarizing soundscape attributes.
In Fig.~\ref{fig:SSQP}, the horizontal axis matches ``Pleasantness,'' and the vertical axis matches ``Eventfulness.''
Other skewed axes (dotted lines) influence to the horizontal and vertical axes by orthogonal projection.

In this study, we call these two dimensions ``soundscape impressions.''
Equation~\ref{equ:PE} calculates the Pleasantness ($P$) and Eventfulness ($E$):
\begin{align}
  P &= (Pl-An) + \cos\left(\frac{\pi}{4}\right) \times \left( Ca-Ch + Vi-Mo \right)%
   \notag\\
  E &= (Ev-Un) + \cos\left(\frac{\pi}{4}\right) \times \left( Ch-Ca + Vi-Mo \right)%
   \label{equ:PE}
\end{align}
Note that the values, $P$ and $E$, are normalized between $-1$ and $+1$.

\subsection{Modification of SSQP}

In this paper, we converted the 5-point scale to 7-point scale for obtaining more accurate values than the original one
\footnote{Normalization factor for Eq.~\ref{equ:PE} is introduced as $6+\sqrt{72}$ for the 7-point scale.}.
Moreover, we introduce two Japanese terms for each soundscape attribute.

The soundscape attributes have translation problems~\cite{Nagahata18_Euronoise,Aletta20_Internoise} that are briefly summarized as the difficulty of one-to-one translation from English to another language.
In Japanese, Nagahata has reported several studies about this problem  \cite{Nagahata18_Euronoise,nagahata_IN19,nagahata_ASJ19en}.
For this reason, we assign two Japanese words to each soundscape attributes as shown in Table~\ref{tbl:insho} referring the Nagahata's works.


\begin{table}[tb]
  \begin{center}
  \caption{Soundscape attributes and their translations in Japanese}
  \label{tbl:insho}
    \begin{tabular}{ll} 
      \hline
      \textbf{Pl}easant   & ta-no-shi:, ko-ko-chi-yo-i \\
                          & \ (\ja{楽しい，心地よい})     \\
      \textbf{Ev}entful   & de-ki-go-to-ga-o:-i, ni-gi-ya-ka-na \\
                          & \ (\ja{出来事が多い，賑やかな}) \\
      \textbf{C}alm       & o-chi-tsu-i-ta, shi-zu-ka-na       \\
                          & \ (\ja{落ち着いた，静かな}) \\
      \textbf{Vi}brant    & ka-Q-ki-ga-a-ru, wa-ku-wa-ku-sa-se-ru \\
                          & \ (\ja{活気がある，ワクワクさせる}) \\
      \hline
      \textbf{An}noying   & so:-zo:-shi:, i-ra-i-ra-sa-se-ru \\
                          & \ (\ja{騒々しい，イライラさせる}) \\
      \textbf{Un}eventful & {\small ko-re-to-i-Q-ta-ko-to-ga-na-i, he:-o-n-bu-ji-na}  \\
                          & \ (\ja{これといった事がない，平穏無事な}) \\
      \textbf{Ch}aotic    & mu-chi-tsu-jo-na, za-tsu-ze-N-to-shi-ta \\
                          & \ (\ja{無秩序な，雑然とした}) \\
      \textbf{Mo}notonous & ta-N-cho:-na, ta-i-ku-tsu-na \\
                          & \ (\ja{単調な，退屈な}) \\
      \hline
    \end{tabular}
  \end{center}
\end{table}

\subsection{Sound-source information in soundscape}

Sound-source information is an important information that affects the evaluations of soundscape attributes~\cite{axelsson10}.
For example, there is a positive correlation between ``Pleasantness'' and ``sounds of nature'', e.g. chirping of the birds, and a negative correlation with ``sounds of technology'', e.g. the sound produced by cars while driving.

Example classification of sound-source information was summarized in the ISO/TS~12913-2~\cite{ISO12913_2}.
The TS indicates the primary three types of sound sources: ``Sounds of technology,'' ``Sounds of nature'' and, ``Sounds of human beings.''
It also demonstrates some examples of the scale, containing a split version for ``Sounds of technology'' into ``Traffic noise'' and ``Other noise.''
Each type of sound source was evaluated whether audible or not on a five-point scale.

In this research, we define seven classes for sound sources as in Table~\ref{tbl:ongen}.
The environmental sound data employed in this research was recorded using smartphones and tablet-type devices \cite{hara_APSIPA17}.
Thus, the sound may be containing recording noise; e.g. tapping noise of the recording terminal, wind noise around the device's microphone, etc.
Such sounds are not present in the actual sound environment and are unsuitable for evaluation as soundscapes.
Thus, we prepare the ``$S_7$: Noise'' indication.

\begin{table}[tb]
  \begin{center}
  \caption{Seven classes for sound-source features}
  \label{tbl:ongen}
    \begin{tabular}{ll}
      \hline
        $S_1$: Technology-traffic & $S_2$: Technology-others \\
        $S_3$: Human-voice        & $S_4$: Human-others \\
        $S_5$: Nature-creature    & $S_6$: Nature-others \\
        $S_7$: Noise & \\
      \hline
    \end{tabular}
  \end{center}
\end{table}

\section{Predictor for soundscape impression}

\begin{figure}[tb]
\centering
    \includegraphics[width=.98\columnwidth]{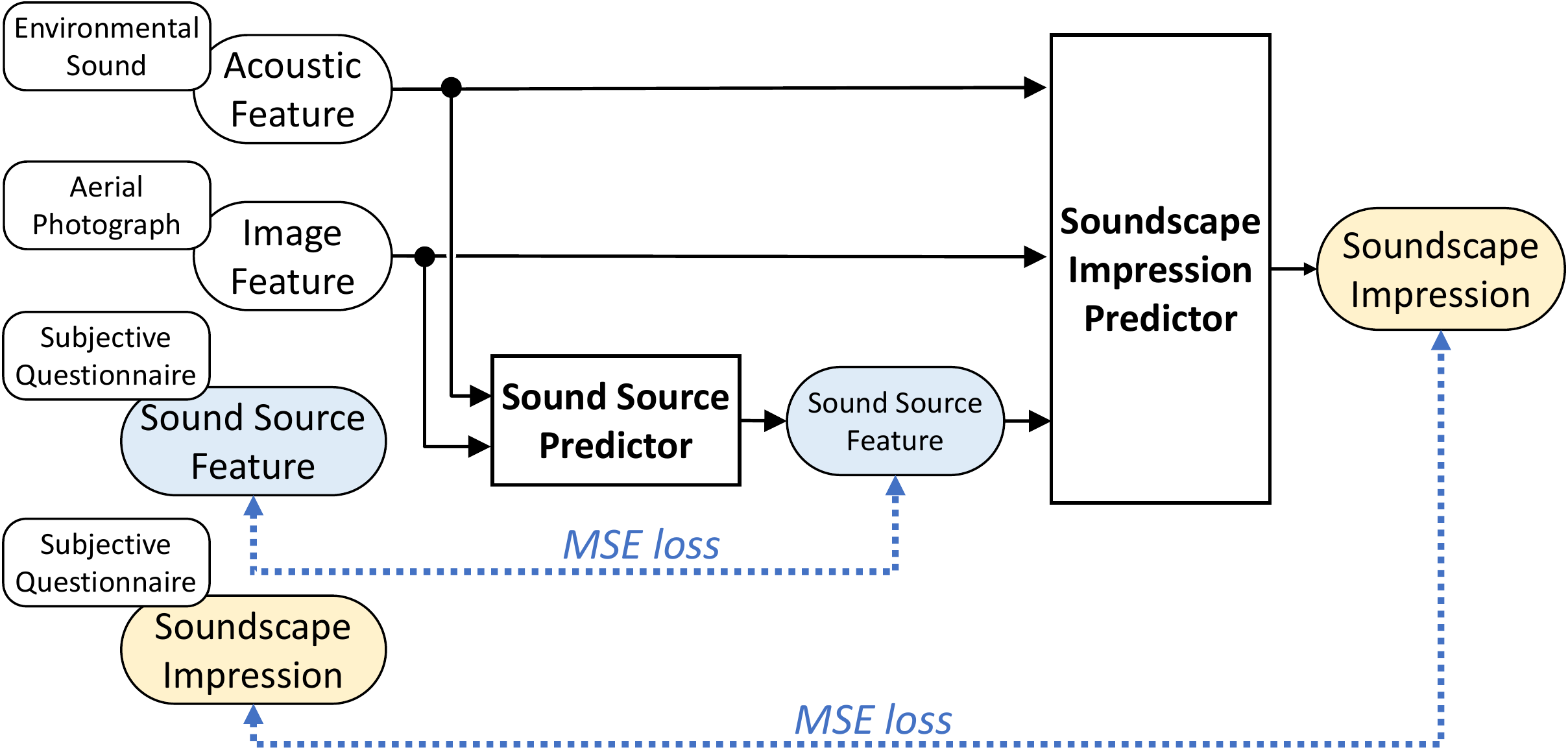}
    \caption{Flowchart of proposed method using predicted sound-source features}
    \label{fig:flowchart2}
\end{figure}

In this study, we propose a two-stage architecture for predicting soundscape impressions.
Figure~\ref{fig:flowchart2} depicts a flowchart of the proposed method.
First, sound-source features are predicted using acoustic features and/or image features.
Then, soundscape impressions are predicted using acoustic features, image features, and/or sound-source features.
True sound-source features and true environment impressions are assigned by humans using questionnaires described in Chapter~\ref{chap:SSImpression}.
Sound-source and soundscape impression predictors are trained using typical machine learning techniques.

Our motivation is to develop predicting method for human assessment values from the acoustic feature and its meta information.
The sound is recorded with meta information, meaning datetime and location by Global Positioning System (GPS).
The acoustic feature is extracted from environmental sounds.
The image feature is generated from an aerial photograph around the recorded location.
Needless to say that the aerial photograph is not an in-situ photograph at recording.
However, it may contain useful information to predict soundscape impressions.
In this modeling, we hope the sound-source predictor is worked as a kind of feature embedding to catch characteristics in various viewpoints for the impressions' prediction than raw acoustic and image features.

\subsection{Acoustic features from Environmental sound}\label{sec:ES}

An acoustic feature is extracted from 10-second sound data.
The feature comprise 126-dimensional values, which is $9 \times (10+4)$.

Each datum comprises equivalent noise levels ($L_{Aeq}$) and outputs of octave-band filters (62.5~Hz, 125~Hz, 250~Hz, 500~Hz, 1000~Hz, 2000~Hz, 4000~Hz, 8000~Hz) for every second for 10~s.
We also used four statistics for these values; i.e., mean, 10\%-tile, 50\%-tile, and 90\%-tile.
Note that each datum is indexed using datetime and location quantified by Quadkey.

\subsection{Image features from Aerial photographs}\label{sec:AP}

\begin{figure}[tb]
	\begin{center}
    \subfloat[][Residential area]{%
    \includegraphics[clip, width=.3\columnwidth]{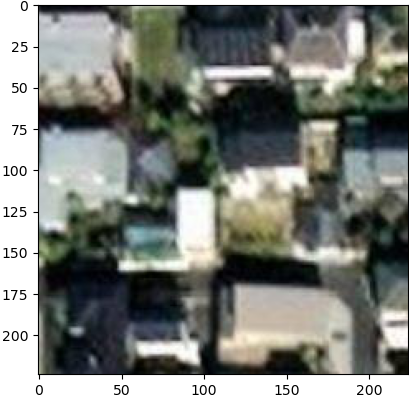}}%
    \hfill
    \subfloat[][Downtown area]{%
    \includegraphics[clip, width=.3\columnwidth]{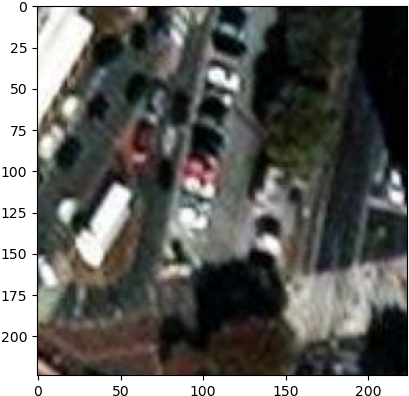}}%
    \hfill
    \subfloat[][Sightseeing area]{%
    \includegraphics[clip, width=.3\columnwidth]{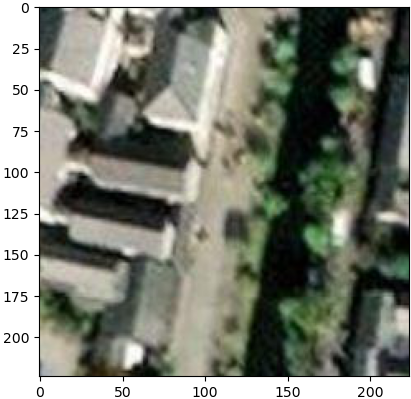}}%
    \hfill
    \caption{Examples of aerial photographs}
    \label{fig3:aerial}
  \end{center}
\end{figure}

Aerial photographs are generated from Bing Map APIs with the location.
Figure~\ref{fig3:aerial} depicts examples of aerial photographs.
We query the photographs as Map tiles using the API%
\footnote{Bing Maps Tile System: \url{https://docs.microsoft.com/en-us/bingmaps/articles/bing-maps-tile-system}}.
Each image is ready as $224\times224$ pixels with zoom level 20 of Bing Map.
The images are made by reform of concatenating and cropping from original images.

\begin{figure}[tb]
\centering
    \includegraphics[clip, width=.9\columnwidth]{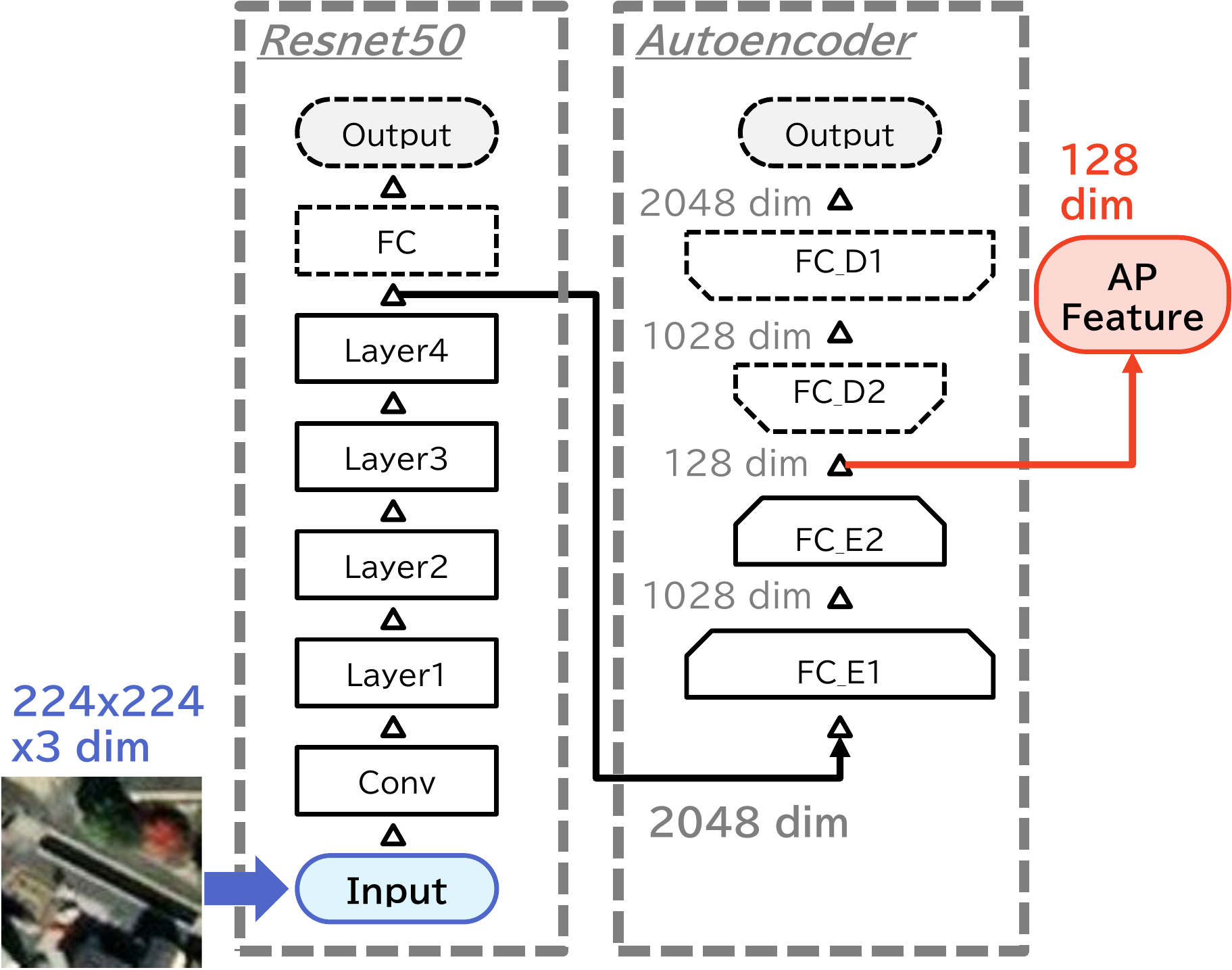}
    \caption{Image feature extractor from aerial photographs}
    \label{fig:ap_extractor}
\end{figure}

In this study, we extracted 128-dimension bottleneck features from each aerial photograph.
Figure~\ref{fig:ap_extractor} shows the image-feature extractor.

First, ResNet-50 converts the aerial photographs to embedded image features.
Specifically, the 2048-dimension output from the layer before the final layer of ResNet-50 is used.

Next, the 2048-dimension features are shrunk using an autoencoder network.
The autoencoder is trained by the 2048-dimension features given as input and output of the network.
It comprises three hidden layers with 1028, 128, and 1028 units, respectively.
Table~\ref{tbl3:AE_joken} shows the autoencoder's training conditions.
We can finally obtain 128-dimension bottleneck features from outputs of the hidden layer.

\begin{table}[tb]
	\begin{center}
	\caption{Training condition of autoencoder for aerial photographs}
	\label{tbl3:AE_joken}
		\begin{tabular}{l|l} 
      \hline
			\# of Epochs            & 100 \\ 
			Batch size          & 16 \\ 
			Activation function & tanh \\
			Loss function       & MSE \\ 
			Optimizer           & SGD \\ 
			- Learning rate     & 0.01 \\ 
			- Momentum          & 0.9 \\
			\# of Data          & 619 \\ 
			Ratio for training data & 90\% \\ 
      \hline
		\end{tabular}
	\end{center}
\end{table}

\subsection{Soundscape impression predictor}

The environmental sounds, aerial photographs, and sound-source features were employed as input for a prediction model of soundscape impressions.
The impressions are the value obtained from two-axes model in Eq.~\ref{equ:PE}.
The model is trained as a DNN model with a multi-layer perceptron model.
The DNN is implemented with python 3.8 and scikit-learn 0.23.1.

We evaluate the model's accuracy by $R^2$ coefficients of determination.
The $R^2$ coefficient of determination is computed by the following Eq.~\ref{equ:r2}:

\begin{equation}
  R^2=1-\frac{\sum_i(y_i-\hat{y_i})^2}{\sum_i(y_i-\bar{y})^2}, \label{equ:r2}
\end{equation}
where $\hat{y}_i$ denotes a predicted value of the impression for $i$-th example and $\bar{y}$ denotes a mean of the impression.

\section{Annotation of SSQP}\label{sec:annotation_ssqp}

\begin{figure}[tb]
\centering
    \includegraphics[clip, width=.9\columnwidth]{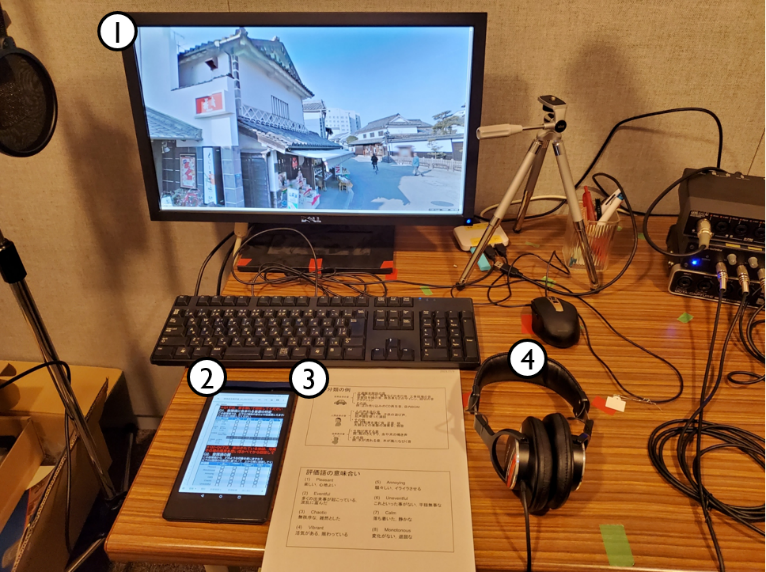}
    \caption{Annotation in a soundproof room. %
             (1) Google street view, %
             (2) Annotation form, %
             (3) Guideline of annotation, %
             (4) Listening to repeatedly playing environmental sound using the headphone}
    \label{fig:bouonshitsu}
\end{figure}

In this study, we prepared the SSQP annotations by one participant.
Among the recorded environmental sound data, 904 environmental sound data were selected for labeling of soundscape attributes and sound-source features with less bias in the recording location and date.
Figure~\ref{fig:bouonshitsu} shows the environmental sounds that were listened to in a soundproof room (background noise level: $20$~dB) using Sony CD900ST headphones.

The participant was shown a Google street view of the area around the recording location during the annotation.
The view was automatically switched to the north, west, south, and east angles every 2.5~s.
At the same time, a 10-second environmental sound was repeatedly played until the participant ended to answer.
The participant answered four questions described below.
Note that, the items of Q1 and Q2 were shuffled to prevent the impact of representing the order.

\noindent\textbf{Q1: Sound source identification}\\\indent
How well do you hear the following seven types of sounds in the surrounding sound environment?
(1: Not at all, 2: A little, 3: Moderately, 4: A lot, 5: Dominates completely;
 Select for seven items shown in Table~\ref{tbl:ongen}.)

\noindent\textbf{Q2: Perceived affecting quality}\\\indent
What is your impression of this sound environment from the viewpoint of these adjectives?
(1: Strongly disagree, $\cdots$, 4: Neither agree nor disagree, $\cdots$, 7: Strongly agree;
 Select for eight items depicted in Table~\ref{tbl:insho}.)

\noindent\textbf{Q3: Assessment of surrounding sound environment}\\\indent
Overall, how would you evaluate the surrounding sound environment?
(1: Very bad, $\cdots$, 4: Neither good nor bad, $\cdots$, 7: Very good;
 Select for 1 item.)

\noindent\textbf{Q4: Assessment of appropriateness}\\\indent
Overall, how do you think the sound environment is suitable for the current location?
(1: Not at all, $\cdots$, 3: Slightly inappropriate, 4: Slightly appropriate, $\cdots$, Perfectly;
 Select for 1 item)

We removed noise-only sounds to cleanse the data.
Thus, we obtained 799 data from 904 environmental sound data.
Each data contains audio data, GPS location (latitude and longitude), and answers for Q1--4.
Note that we use answers of Q1 and Q2 for the following experiments.

\section{Evaluation experiments}

We confirm what feature is crucial for the estimation of impression information by comparing the accuracy of the proposed method with various patterns of input features like acoustic features and aerial photographs.
First, we compare the prediction accuracy of the impression predictor built by combining environmental sounds, aerial photographs, and oracle sound-source features.
Then, we compare the prediction accuracy of the impression predictor using environmental sounds, aerial photographs, and predicted sound-source features.

\subsection{Experimental data}

In this investigation, we used 799 data annotated in Section~\ref{sec:annotation_ssqp}.
The environmental sound data were recorded between 2014 and 2017 \cite{hara_ICME16,hara_APSIPA17} at around of Okayama University, Okayama Station, and the Kurashiki Bikan Historical Park, which was corresponding to example photographs in Fig.~\ref{fig3:aerial} (a)--(c), respectively.
The environmental sound data were recorded using an application running on an Android tablet, Google Nexus 7.
Sounds are recorded at a sampling frequency of 32~kHz and 16~bits over a single channel.

Section~\ref{sec:ES} describes the extracted acoustic features from the audio data.
The data contains location information, thus, we can generate aerial photographs using Bing Maps described in Section~\ref{sec:AP}.
After processing these images, 619 unique aerial photographs were finally prepared to cover the recorded locations.

\subsection{Experiment for soundscape impression predictor using oracle sound-source features}

In this section, we build impression predictors with environmental sound, aerial photographs, and sound-source features, that are gathered by previous Section~\ref{sec:annotation_ssqp}.
The following six patterns of feature are prepared to compare the significance of each feature for the prediction.

\begin{itemize}
  \item \textsf{ES} environmental sound (126~dim.)
  \item \textsf{ES+SS} environmental sound and sound-source features (133~dim.)
  \item \textsf{AP} aerial photograph (128~dim.)
  \item \textsf{AP+SS} aerial photograph and sound-source features (135~dim.)
  \item \textsf{ES+AP} environmental sound and aerial photograph (254~dim.)
  \item \textsf{ES+AP+SS} environmental sound, aerial photograph and sound-source features (261~dim.)
\end{itemize}

Table~\ref{tbl3:DNN_joken} shows the training condition of the DNN model.
The models are built for each pattern of feature, thus, the optimal hyperparameters of the models are different from the features.
\begin{table}[tb]\centering
  \caption{Training condition of impression predictor}
  \label{tbl3:DNN_joken}
  \begin{tabular}{l|l}
  \hline
  \textbf{Estimator} & \\
  \ Base model & DNN (Multi-layer Perceptron)           \\
  \ \#Data     & 799 (training: 599, test: 200)         \\
  \ Regularization & $10^{-3}$ \\
  \ Activation function & ReLU \\
  \ Optimizer & Adam \\
  \hline
  \textbf{Parameter search}  & \\
  \ Target parameters & \\
  \ - \# of hidden layers  & $1, 2, \cdots, 10$         \\
  \ - \# of units          & $2^2, 2^3, \cdots, 2^{10}$ \\ 
  \ Search algorithm & Treestructured Parzen Estimator: TPE \\
                     & (Optuna ver 2.10.0)              \\
  \ \# of iteration  & 100                              \\
  \ Parameter selection & Maximization of $R^2$ coefficient \\
                        & with 10-fold cross validation \\
  \hline
  \end{tabular}
\end{table}

\begin{figure}[tb]
  \begin{center}
    \includegraphics[width=.88\columnwidth]{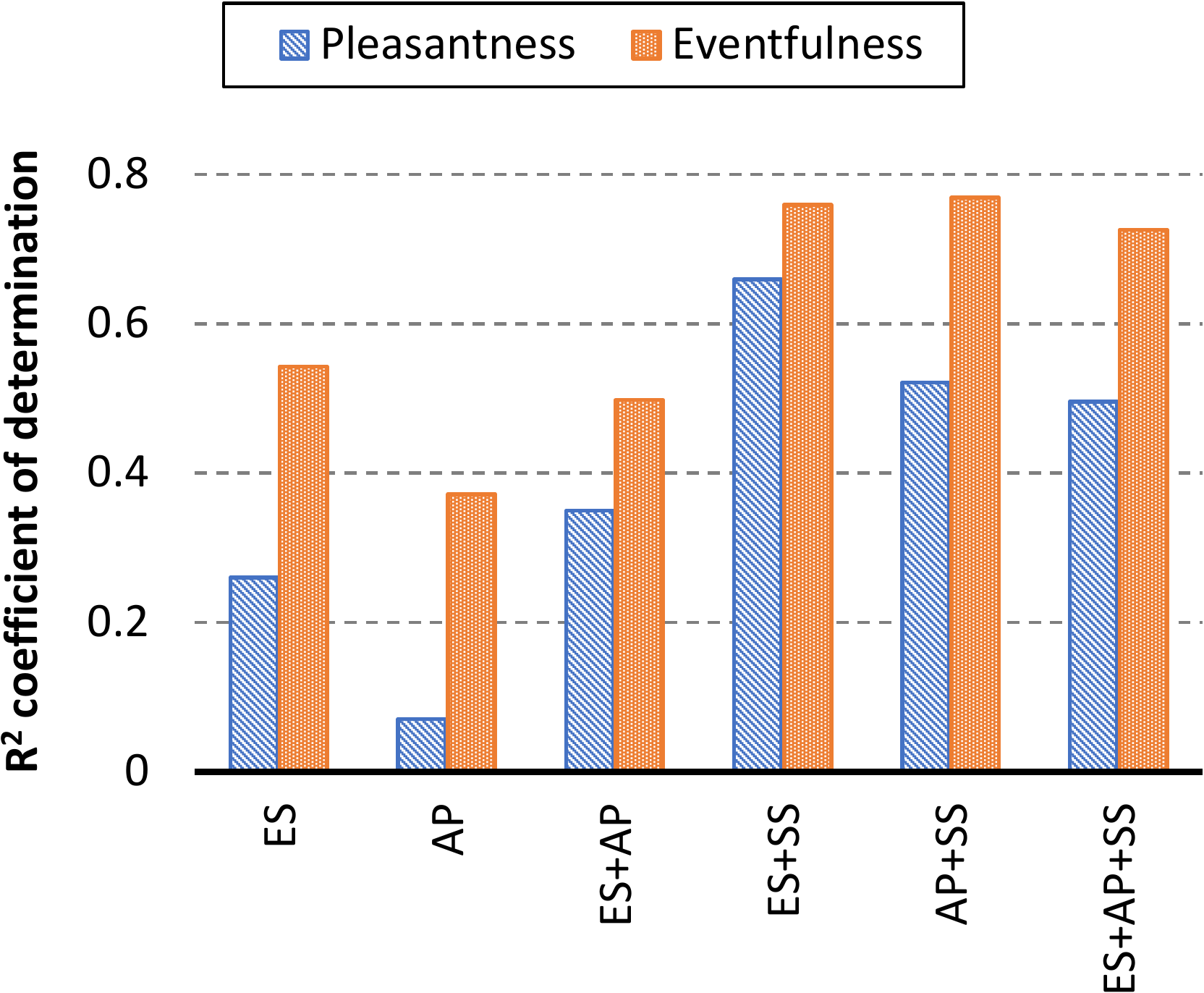}
    \caption{Result of impression predictor with oracle sound-source features}
    \label{fig:6pt_dnn}
  \end{center}
\end{figure}

Figure~\ref{fig:6pt_dnn} shows the accuracy of the impression predictors.
The highest accuracy of Pleasantness was attained using \textsf{ES+SS}, with an $R^2$ coefficient of determination of 0.659.
The highest accuracy of Eventfulness was attained using \textsf{AP+SS}, with an $R^2$ coefficient of determination of 0.769.

For Pleasantness, the prediction accuracy is the highest when using both environmental sounds and sound-source features \textsf{ES+SS}.
This is believed that the sound level and types have a substantial influence on the soundscape attributes, meaning Pleasant and Annoying.

For Eventfulness, the prediction accuracy is the highest when using both aerial photographs and sound-source features (\textsf{AP+SS}).
This is believed that the attributes, that is Eventful and Uneventful, are highly influenced by the location's atmosphere.
Thus, aerial photographs that can capture location-based characteristics might be suitable.

\subsection{Experiment for soundscape impression predictor using predicted sound-source features}

We consider using the predicted values for the sound-source features since the sound-source features are manually assigned.
The sound-source predictor may use environmental sounds, aerial photographs, or both environmental sounds and aerial photographs as input features.
Thus, the following three patterns of sound-source feature usage are feasible.

\begin{itemize}
  \item \textsf{eSS[ES]} sound-source features predicted with environmental sound
  \item \textsf{eSS[AP]} sound-source features predicted with an aerial photograph
  \item \textsf{eSS[ES+AP]} sound-source features predicted with environmental sound and aerial photograph
\end{itemize}

Combined with the three input patterns of the impression predictor described in the previous section, learning using $3\times3=9$ pattern features.




\begin{figure*}[t]
    \centering
    \includegraphics[width=.8\textwidth]{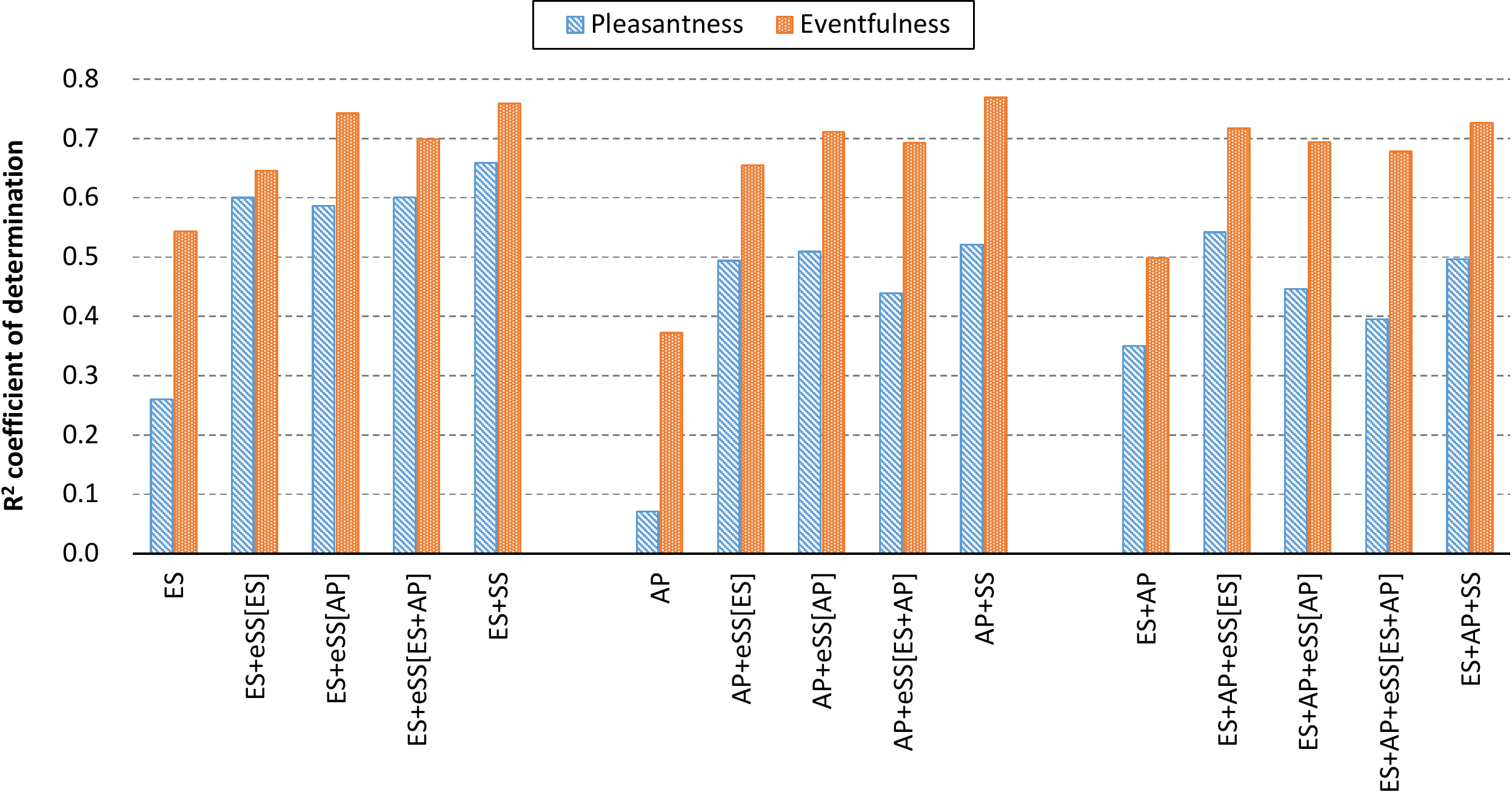}%
    \caption{Accuracy of sound-source estimator using environmental sound, aerial photographs, and sound-source features}
    \label{fig:9pt}
\end{figure*}


Figure~\ref{fig:9pt} shows the prediction results.
The model with the highest accuracy of ``Pleasantness'' was \textsf{ES+eSS[ES+AP]} with the $R^2$ coefficient of determination of 0.601.
The model with the highest accuracy of ``Eventfulness'' was \textsf{ES+eSS[AP]} with the $R^2$ coefficient of determination of 0.742.

The accuracy of ``Pleasantness'' and ``Eventfulness'' are only 0.06 and 0.03 lower than the oracle models,
compared with the result of \textsf{ES+SS} given the oracle sound-source feature, respectively.
The result suggests that impression predictors could be realized without manually annotated labels.

\section{Conclusions}

In this study, we proposed a method for predicting soundscape impressions using environmental sound and aerial photographs.
First, we experimented by comparing the accuracy of the impression predictor using a total of six patterns of features.
The results suggest that the use of environmental sounds for Pleasantness and aerial photographs for Eventfulness enhances the accuracy.
Then, we confirmed the effectiveness of predicted sound-source features by comparing the oracle sound-source features.
The experimental results shows the degradation of accuracy was adequately low, and it suggests the effectiveness of using predicted sound-source features to predict the soundscape impressions.

We showed that our approach has the potential to predict soundscape impressions with acoustic data and its location information.
However, there are remaining numerous future works.
First, we only showed the results by annotating one person, thus, we should employ more annotators and confirm re-productivity for our proposed method.
Second, while the experimental data were mixed with two cities, meaning Okayama and Kurashiki, we should experiment focused on the difference in characteristics of the cities.
Finally, a development of a joint-training method for two predictors in our model is also important future work.



\bibliographystyle{IEEEtran}
\bibliography{apsipa2022_ono}

\begin{thebibliography}{10}
\providecommand{\url}[1]{#1}
\csname url@samestyle\endcsname
\providecommand{\newblock}{\relax}
\providecommand{\bibinfo}[2]{#2}
\providecommand{\BIBentrySTDinterwordspacing}{\spaceskip=0pt\relax}
\providecommand{\BIBentryALTinterwordstretchfactor}{4}
\providecommand{\BIBentryALTinterwordspacing}{\spaceskip=\fontdimen2\font plus
\BIBentryALTinterwordstretchfactor\fontdimen3\font minus
  \fontdimen4\font\relax}
\providecommand{\BIBforeignlanguage}[2]{{%
\expandafter\ifx\csname l@#1\endcsname\relax
\typeout{** WARNING: IEEEtran.bst: No hyphenation pattern has been}%
\typeout{** loaded for the language `#1'. Using the pattern for}%
\typeout{** the default language instead.}%
\else
\language=\csname l@#1\endcsname
\fi
#2}}
\providecommand{\BIBdecl}{\relax}
\BIBdecl

\bibitem{ISO12913_1}
{ISO 12913-1:2014(E)}, ``{Acoustics} --{Soundscape}-- {Part} 1: Definition and
  conceptual framework,'' International Organization for Standardization,
  Geneva, CH, Standard, Sep. 2014.

\bibitem{axelsson10}
{\"O}.~Axelsson, M.~E. Nilsson, and B.~Berglund, ``A principal components model
  of soundscape perception,'' \emph{The Journal of the Acoustical Society of
  America}, vol. 128, no.~5, pp. 2836--2846, 2010.

\bibitem{lunden2016urban}
P.~Lund{\'e}n and M.~Hurtig, ``On urban soundscape mapping: A computer can
  predict the outcome of soundscape assessments,'' in \emph{Proceedings of
  INTER-NOISE 2016}, no.~6, 2016, pp. 2017--2024.

\bibitem{pijanowski2011soundscape}
B.~C. Pijanowski, L.~J. Villanueva-Rivera, S.~L. Dumyahn, A.~Farina, B.~L.
  Krause, B.~M. Napoletano, S.~H. Gage, and N.~Pieretti, ``Soundscape ecology:
  the science of sound in the landscape,'' \emph{BioScience}, vol.~61, no.~3,
  pp. 203--216, 2011.

\bibitem{axelsson15}
{\"O}.~Axelsson, ``How to measure soundscape quality,'' in \emph{Proceedings of
  the Euronoise 2015 conference}, 2015, pp. 1477--1481.

\bibitem{ISO12913_2}
{ISO/TS 12913-2:2018(E)}, ``{Acoustics} --{Soundscape}-- {Part} 2: Data
  collection and reporting requirements,'' International Organization for
  Standardization, Geneva, CH, Standard, Aug. 2018.

\bibitem{Rana2010_IPSN_EarPhone}
R.~Rana, C.~Chou, S.~Kanhere, N.~Bulusu, and W.~Hu, ``{Ear}-{Phone}: An
  end-to-end participatory urban noise mapping system,'' in \emph{Proceedings
  of IPSN-2010}, Apr. 2010, pp. 105--116.

\bibitem{Kanjo2010_NoiseSPY}
E.~Kanjo, ``{NoiseSPY}: A real-time mobile phone platform for urban noise
  monitoring and mapping,'' \emph{Mobile Networks and Applications}, vol.~15,
  no.~4, pp. 562--574, Aug. 2010.

\bibitem{DHondt2012_Journal_NoiseTube}
E.~D'Hondt, M.~A. Stevens, and A.~Jacobs, ``Participatory noise mapping works!
  an evaluation of participatory sensing as an alternative to standard
  techniques for environmental monitoring,'' \emph{Pervasive and Mobile
  Computing}, vol.~9, no.~5, pp. 681--694, Oct. 2013.

\bibitem{Mydlarz_InterNoise2011}
C.~Mydlarz, I.~Drumm, and T.~Cox, ``Application of novel techniques for the
  investigation of human relationships with soundscapes,'' in \emph{Proceedings
  of INTERNOISE 2011 congress}, Sep. 2011, pp. 738--744.

\bibitem{hara_ICME16}
S.~Hara, S.~Kobayashi, and M.~Abe, ``Sound collection systems using a
  crowdsourcing approach to construct sound map based on subjective
  evaluation,'' in \emph{Proceedings of ICME Workshop}, 2016, pp. 1--6.

\bibitem{hara_APSIPA17}
S.~Hara, A.~Hatakeyama, S.~Kobayashi, and M.~Abe, ``Sound sensing using
  smartphones as a crowdsourcing approach,'' in \emph{Proceedings of APSIPA-ASC
  2017}, 2017, pp. 1328--1333.

\bibitem{ISO12913_3}
{ISO/TS 12913-3:2019(E)}, ``{Acoustics} --{Soundscape}-- {Part} 3: Data
  analysis,'' International Organization for Standardization, Geneva, CH,
  Standard, Dec. 2019.

\bibitem{Nagahata18_Euronoise}
K.~Nagahata, ``Linguistic issues we must resolve before the standardization of
  soundscape research,'' in \emph{Proceedings of EURO-NOISE 2018}, 2018, p.
  2459^^e2^^80^^932464.

\bibitem{Aletta20_Internoise}
F.~Aletta, T.~Oberman, {\"O}.~Axelsson, and et~al., ``Soundscape assessment:
  towards a validated translation of perceptual attributes in different
  languages,'' in \emph{Proceedings of INTER-NOISE 2020}, Aug. 2020.

\bibitem{nagahata_IN19}
K.~Nagahata, ``Examination of soundscape-quality protocols in japanese,'' in
  \emph{Proceedings of INTER-NOISE 2019}, no.~9, 2019, pp. 437--446.

\bibitem{nagahata_ASJ19en}
K.~{Nagahata}, ``A research method of sonic environment: The research method
  for soundscape studies provided by the {ISO} 12913 series (in {Japanese}),''
  \emph{The Journal of the Acoustical Society of Japan}, vol.~75, no.~8, pp.
  473--480, 2019.

\end{thebibliography}

\end{document}